%% file: main.tex
\documentclass[journal]{IEEEtran}

\usepackage[utf8]{inputenc} 
\usepackage[T1]{fontenc}    
\usepackage{hyperref}       
\usepackage{url}            
\usepackage{booktabs}       
\usepackage{amsfonts}       
\usepackage{nicefrac}       
\usepackage{microtype}      
\usepackage{lipsum}
\usepackage{graphicx}
\graphicspath{ {./images/} }

\usepackage{multicol}
\usepackage[noend]{algpseudocode}
\usepackage{ragged2e}
\usepackage{amsmath}
\usepackage{orcidlink}
\usepackage{ebproof}
\usepackage{color,soulutf8}
\sethlcolor{lightgray}
\usepackage{syntax}
\usepackage{calc}
\usepackage{listings}
\usepackage{url}
\usepackage{booktabs}
\usepackage{multirow}
\usepackage{pdfpages}
\usepackage{subcaption}
\usepackage[font=small]{caption}
\usepackage{proof}
\usepackage[linesnumbered,noend]{algorithm2e}
\SetKwInput{Type}{Type} 
\usepackage{algpseudocode}
\usepackage[export]{adjustbox}

\usepackage{booktabs,amsfonts,dcolumn}
\newcommand{\tool}{SpotOn}
\usepackage{amsmath,amsfonts}
\usepackage{lipsum}
\usepackage{textcomp}
\newcolumntype{d}[1]{D..{#1}}
\usepackage{booktabs}
\newcommand*\xbar[1]{%
  \hbox{%
    \vbox{%
      \hrule height 0.5pt 
      \kern0.5ex
      \hbox{%
        \kern-0.1em
        \ensuremath{#1}%
        \kern-0.1em
      }%
    }%
  }%
} 
\usepackage{lipsum}
\usepackage{wrapfig}

\usepackage{etoolbox}
\newcommand*\Suppressnumber{%
  \lst@AddToHook{OnNewLine}{%
    \let\thelstnumber\relax%
     \advance\c@lstnumber-\@ne\relax%
    }%
}

\newcommand{\throwdef}[1]{{\bf throw}}

\newif\ifreview
\usepackage{etoolbox}

\newcommand{\repthanks}[1]{\textsuperscript{\ref{#1}}}
\makeatletter
\patchcmd{\maketitle}
  {\def\thanks}
  {\let\repthanks\repthanksunskip\def\thanks}
  {}{}
\patchcmd{\@maketitle}
  {\def\thanks}
  {\let\repthanks\@gobble\def\thanks}
  {}{}
\newcommand\repthanksunskip[1]{\unskip{}}
\makeatother

\lstdefinestyle{base}{
  moredelim=**[is][\color{blue}]{@}{@},
  moredelim=**[is][\color{red}]{*}{*},
}
\usepackage{amssymb}
\usepackage{hyperref}
\begin{document}

\title{Generator-Based Fuzzers with Type-Based Targeted Mutation}

\author{
 Soha Hussein~\IEEEmembership{Member,~IEEE,},  Stephen McCamant~\IEEEmembership{Member,~IEEE,}, and  Michael W. Whalen~\IEEEmembership{Member,~IEEE,}
 \thanks{S. Hussein with the University of Minnesota and Ain Shams University (soha@umn.edu, soha.hussein@cis.asu.edu.eg)}
 \thanks{S. McCamant with the University of Minnesota (mccamant@cs.umn.edu)} 
 \thanks{M. Whalen with the University of Minnesota, (mwwhalen@umn.edu)}
}

\maketitle
\begin{abstract}

Generator-Based Fuzzers (GBFs) construct inputs using programmable generators, making them effective for applications with complex structured inputs. However, GBFs typically use uniform mutation strategies that treat all input portions equally. As inputs grow larger, the most valuable mutation locations—those affecting control-flow decisions—constitute an increasingly smaller fraction of the input space, limiting fuzzer effectiveness.

Previous approaches use heavyweight techniques such as taint analysis or symbolic execution to identify valuable mutation targets. However, these techniques are impractical for modern applications with cross-language interactions and distributed execution, such as serverless computing.

In this paper, we introduce a type-based mutation heuristic for Java GBFs that leverages lightweight static analysis. Our key insight is that type information, readily available in Java's runtime, can serve as an effective proxy for identifying which input portions influence branching decisions. We statically identify types that flow into branch conditions, dynamically annotate generated inputs with type information, and prioritize mutations of input segments corresponding to influencing types of uncovered branches. 

We implement our technique as an extension to Zest, the state-of-the-art Java GBF, and evaluate it on seven AWS Lambda applications. Our results show a geometric mean improvement of 16\% in application branch coverage and 14\% in total coverage (including third-party libraries), with individual benchmark improvements ranging from 2\% to 61\%. Statistical analysis confirms significant improvements on four of seven benchmarks, demonstrating that type-based targeting provides practical benefits when application branches have clear type dependencies. 
\end{abstract}
\begin{IEEEkeywords}
Testing and Debugging, Program analysis, Testing tools.
\end{IEEEkeywords}


\input{introduction}
\input{motivation}
\input{technique}
\input{evaluation}
\input{related-work}
\input{conclusion}
\bibliographystyle{IEEEtran}  

\bibliography{references}

\end{document}

%% file: introduction.tex
\section{Introduction}
Fuzzing has become an indispensable technique for finding bugs in production software, yet its effectiveness diminishes dramatically as applications grow in complexity and input size. Two techniques, coverage feedback and input generators, help make fuzzing more practical, but we observe a missed opportunity when they are used together for larger inputs and codebases.
Coverage-guided fuzzing (hereafter CGF''), as popularized in AFL~\cite{aflsite}, augments mutational fuzzing by recording which code locations a test input covers. CGF then prioritizes inputs that cover previously-uncovered code (intuitively, interesting inputs'') as the starting points for future mutations. CGF is effective at biasing the search process towards relevant inputs without requiring expensive analysis, and this power has made it popular both for practical bug-finding and as the basis for research on further enhancements~\cite{aflfast, aflgo,tensorfuzz, coverageguidedtracing,aflsmart,Collafl,fuzzphantom}. However, CGF does not imply any change to the mutation process, which is still random.

Another idea that can improve the effectiveness of fuzzing is to capture patterns of relevant input in code with input generators. An input generator for a given type (also commonly used in property-based testing such as QuickCheck~\cite{quickcheck}) is just a function that returns a random instance of the type. Using generators to provide inputs yields generator-based fuzzing (hereafter ``GBF''). GBF can be used for both unit and system testing, and can capture patterns that constitute legal and typical inputs. 
Authors of the Zest system~\cite{zest} demonstrated how CGF and GBF approaches can be combined by applying mutations to the sequence of random choices a generator uses to build an object. Such mutations are more likely to achieve semantically relevant coverage because the generators enforce input validity constraints. 
However, even when CGF and GBF are combined as in Zest, the location for a mutation is still chosen uniformly at random across the input. We observe that such tools still have no connection between their coverage goals and the process of selecting mutations.
This uniform mutation strategy becomes increasingly problematic as both inputs and codebases grow. For larger inputs, the most valuable mutation locations, such as fixed-size metadata fields in image files or protocol headers, shrink as a fraction of total input size. For example, a 100-byte metadata field in a 1MB image represents only 0.01\% of the input space, yet uniform mutation treats it no differently than irrelevant pixel data. Similarly, when fuzzing applications built on extensive libraries and frameworks, the application code itself may constitute only a small fraction of total executed code. We will use the term application'' to refer generically to the often smaller code that was written for a specific use case and that is the primary target of fuzzing. We will use third-party code'' to refer generically to all of the other code that also executes as part of the program under test but is not the primary target of fuzzing, such as libraries and frameworks. If fuzzers cannot distinguish between coverage in application code versus third-party libraries, they waste effort exploring irrelevant library internals while missing critical application branches. This can lead to an illusion of progress, as increased coverage in third-party code does not necessarily translate to meaningful coverage in the application code.


More expensive analysis techniques such as dynamic taint tracking or symbolic execution could be used to trace which parts of the input are relevant to code~\cite{taindirectedWhiteboxfuzzing, driller, confetti}. However, tracking complete data flow in the cross-language and distributed environments found in modern applications, such as serverless computing, is extremely difficult. 

Our key insight is that types provide a lightweight yet effective proxy for identifying which input parts influence application behavior. In languages like Java, the type of an object that flows to a branch condition inherently constrains which input fields could have affected that branch. For instance, if a branch condition checks a Metadata object's timestamp field, we can infer that the input's metadata-generating code is more relevant to covering that branch than code generating unrelated pixel data. Unlike heavyweight techniques requiring complete data-flow tracking, type information is readily available in Java's runtime and can bridge the gap between input structure and program branches without expensive instrumentation.


Our approach links input generation to application code via types through a combination of static and dynamic analysis. Statically, we identify \emph{influencing types}, types whose objects are data-flow predecessors of branch conditions in the program under test. This produces a ranked list of influencing types for each branch, which later guides the fuzzer toward uncovered application branches by prioritizing mutations of input generators responsible for constructing objects of these types. 
Dynamically, we associate types with objects constructed by input generators. Since generator methods commonly call other generator methods to construct complex types, our implementation uses a simple form of execution indexing~\cite{baseEIPaper} to record type information for input-generating code in the context of its call stack.
During fuzzing, our technique matches types statically associated with uncovered branches to types dynamically associated with parts of the input 
, increasing the frequency of mutating input parts corresponding to objects of matched influencing types. As in other forms of CGF, this mutation occurs in a loop that measures whether mutated inputs achieve new code coverage, and saves inputs that trigger new coverage to be a basis for future mutations. As an additional improvement to accelerate the fuzzing process, our static analysis also collects constant strings from the application code and includes these strings as possibilities when generating strings in program inputs.


We implemented this type-based heuristic on top of Zest and evaluated it on a set of AWS Lambda applications available on GitHub.
We use Serverless applications as they present a particular fuzzing challenge: application code—though potentially complex—is often dwarfed by extensive third-party libraries in terms of overall code volume. When application logic makes control-flow decisions based on small portions of structured input while libraries process the bulk of that input, uniform fuzzing strategies waste effort on library code at the expense of application coverage

Our results 
shows that our technique achieves a geometric mean improvement of 18.2\% in application branch coverage over the baseline. When measuring total coverage (including third-party code) across the same set of libraries, we observe a 43.2\% geometric mean improvement, with even larger gains when our approach enables coverage of additional libraries previously unreached by the baseline.
These results demonstrate that type-based targeting provides a practical, lightweight alternative to heavyweight analysis techniques for guiding fuzzer exploration

The paper is organized as follows: Section~\ref{sec:motivationalExample} provides a motivating example, Section~\ref{sec:technique} outlines our technique, Section~\ref{sec:evaluation} presents our evaluation, Section~\ref{sec:related-work} discusses related work, and Section~\ref{sec:conclusion} concludes.

%% file: motivation.tex
\section{Motivating Example}
\label{sec:motivationalExample}

\begin{figure}
    \centering    
\begin{subfigure} {.50\textwidth}
    \includegraphics[width=\columnwidth]{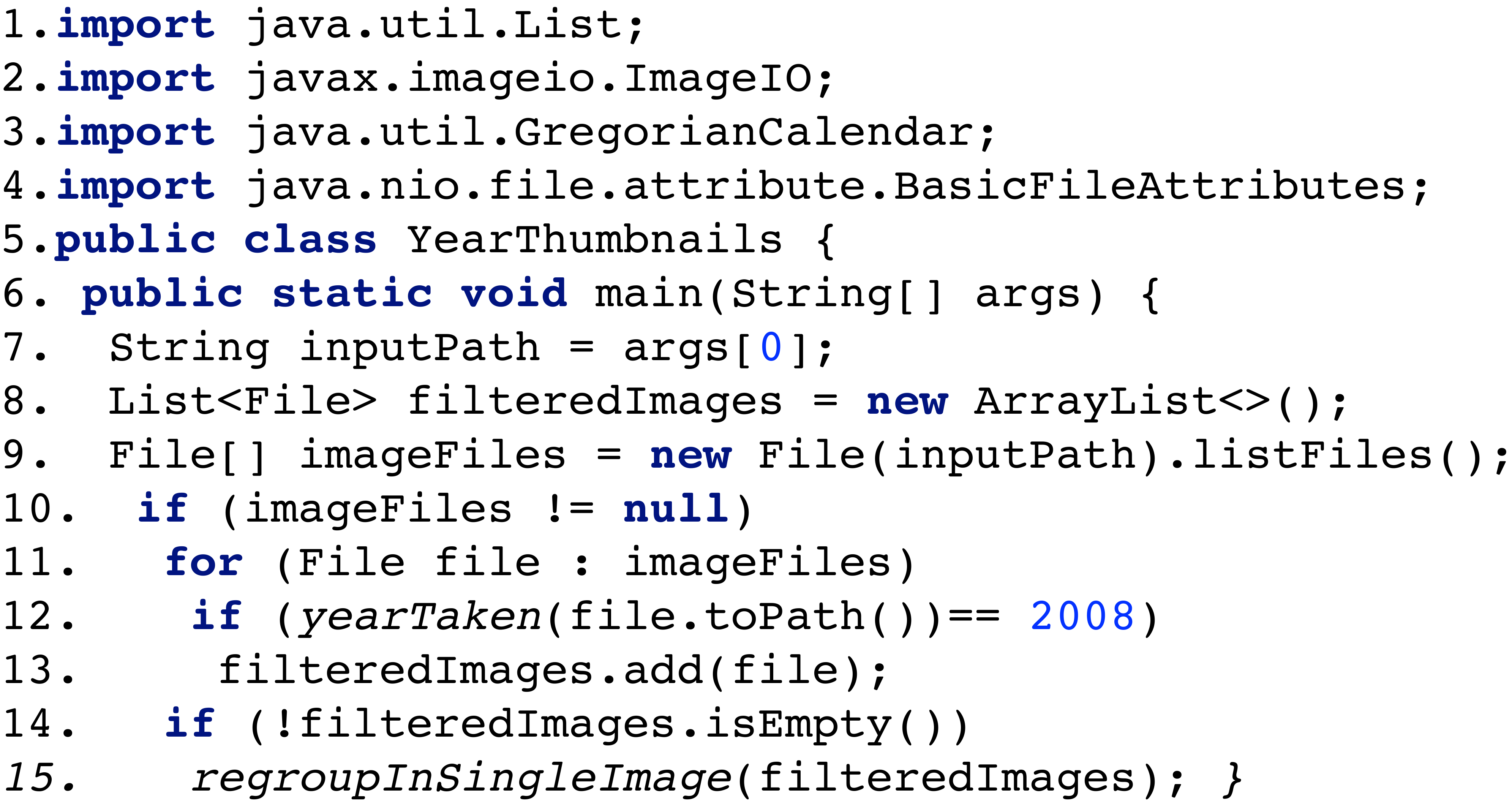}
    \caption{Program for creating thumbnails for images modified in 2008.}
    \label{fig:yearThumbnails}
\end{subfigure}
\begin{subfigure}{.50\textwidth}
    \includegraphics[width=\columnwidth]{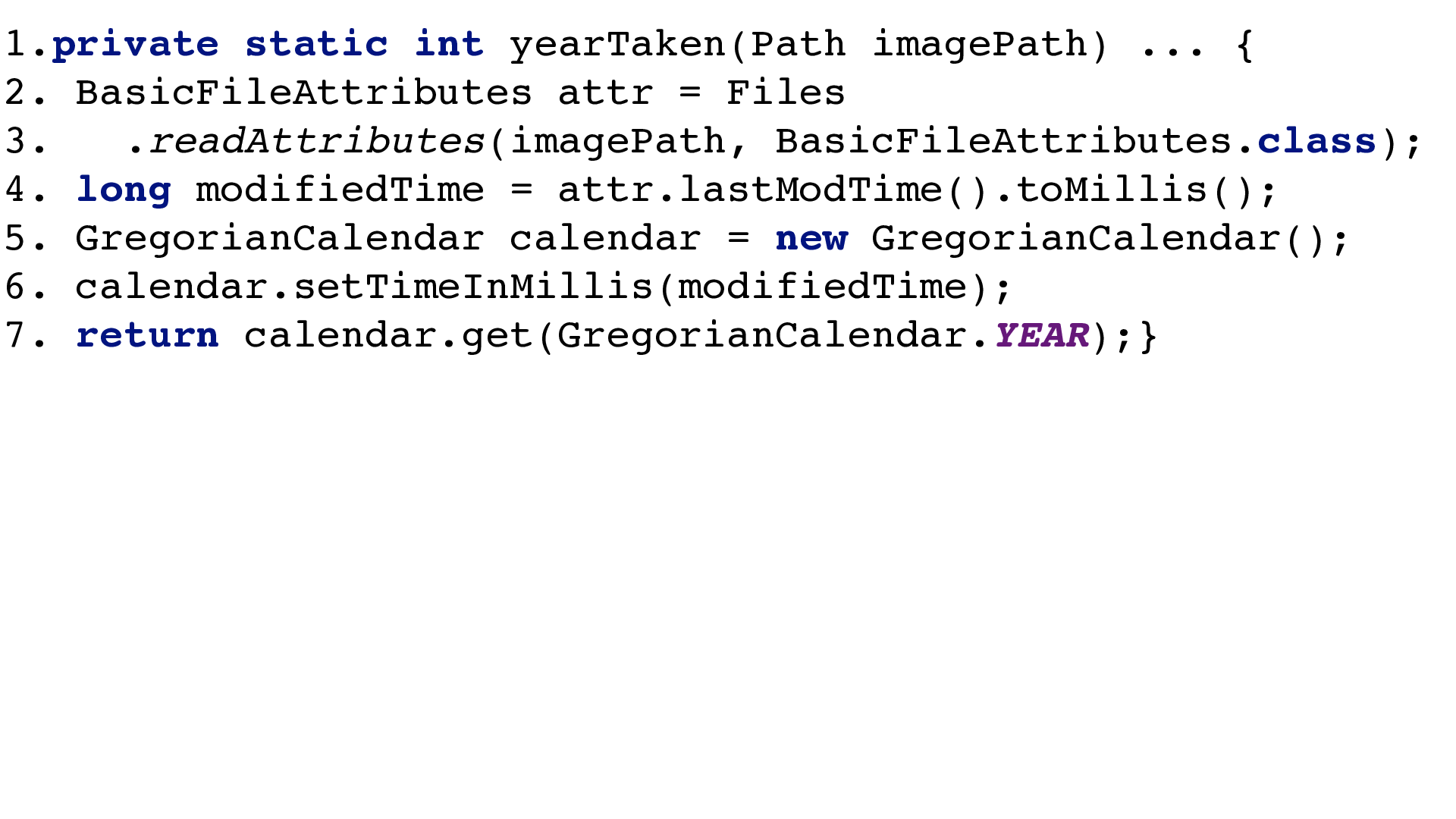}
    \vspace*{-30mm}
    \caption{Helper method for computing the getting modification year.}
    \label{fig:yearTaken}
\end{subfigure}
\caption{A program snippet that creates thumbnails for images of a specific year.}
    \label{fig:motivationalExample}
\end{figure}

We illustrate the fuzzing challenges through a simple yet representative example: an image processing program that filters files by modification year. Despite its simplicity, this program is difficult for uniform mutation strategies to cover effectively.
More precisely, consider the program in Fig.~\ref{fig:yearThumbnails}.
The program creates a single image composed of thumbnails of all images modified in the {\tt year} 2008. Given the location of image files in {\tt inputPath} {\bf(line 7)}, the program first finds all files within the specified path {\bf(line 9)}, then it iterates over all image files within this directory to collect those whose modification date matches the intended year  {\bf(lines 11-13)}. The helper function {\tt yearTaken} (in Fig.~\ref{fig:yearTaken}) finds the system modification year of a file object. 
The method first attempts to read attributes of the file system for the image path, particularly the last modified time {\bf(line 2-4)}. Then, it converts that time into a date form captured in a {\tt GregorianCalendar} object {\bf(line 5-6)}. Finally, the method extracts and returns the year of the last modification of the file  {\bf(line 7)}. 
The thumbnail program in Fig.~\ref{fig:yearThumbnails} creates a combined thumbnail image for the filtered images {\bf(line 15)}, using some image processing happening within the {\tt regroupInSingleImage} method, whose details are omitted. 

To understand why this program is challenging for fuzzers, we first explain how generator-based fuzzers work, then demonstrate how they would approach fuzzing this specific program, then finally present how our heuristic work.

%
%
%


\subsection{Generator-Based Fuzzers Background}
\begin{wrapfigure}{r}{0.60\columnwidth}
    \centering
    \includegraphics[width=0.9\linewidth]{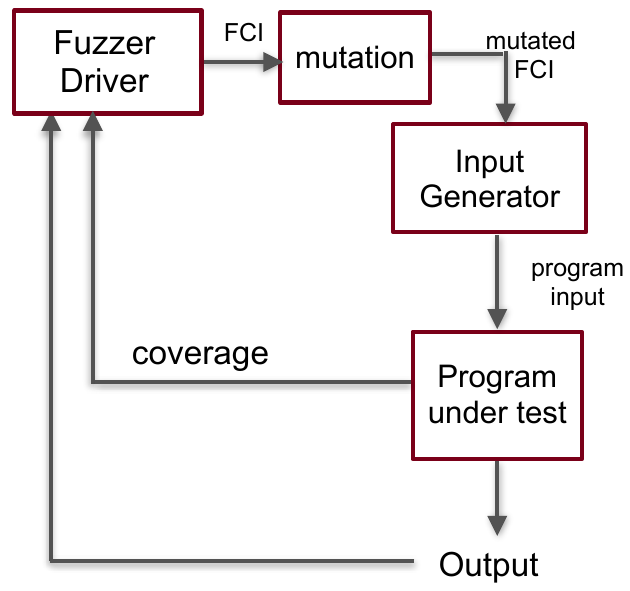}
    \caption{Generator-Based Fuzzer - FCI is the fuzzer chosen input}
    \label{fig:GBF}
\end{wrapfigure}

To fuzz a program, GBF's fuzzing driver works as follows (Fig.\ref{fig:GBF}). First, the fuzzer driver starts by creating program inputs. This step happens indirectly by running the input generators on the Fuzzer-Chosen Input (FCI), see Fig.\ref{fig:GBF}. 
The \textit{FCI} captures the random values used by the input generators to construct a random program input. 
The FCI is consulted whenever the input generator makes a pseudo-random number call. 
This design allows the GBF to replay previous program inputs, by reusing the same FCI sequences, or create new program inputs by mutating subsequences within the FCI; i.e., using the mutation process. 
Next, the program under test is executed on the constructed program input, while collecting coverage and output information. Useful inputs, i.e., inputs that achieved new coverages, are saved act as the base of mutation in subsequent fuzzing.

Program inputs generated by this approach are more likely to explore deeper parts of the program under test than those created by traditional CGF. This is because generated program inputs are structurally correct, enabling them to bypass the program’s well-formedness syntax checks.
However, because it relies on a uniform random mutation strategy to generate new inputs, finding the right combination to trigger interesting program behavior becomes increasingly difficult as the input space increases.

\begin{figure}
    \centering    
\begin{subfigure}{.50\textwidth}
    \includegraphics[width=\columnwidth]{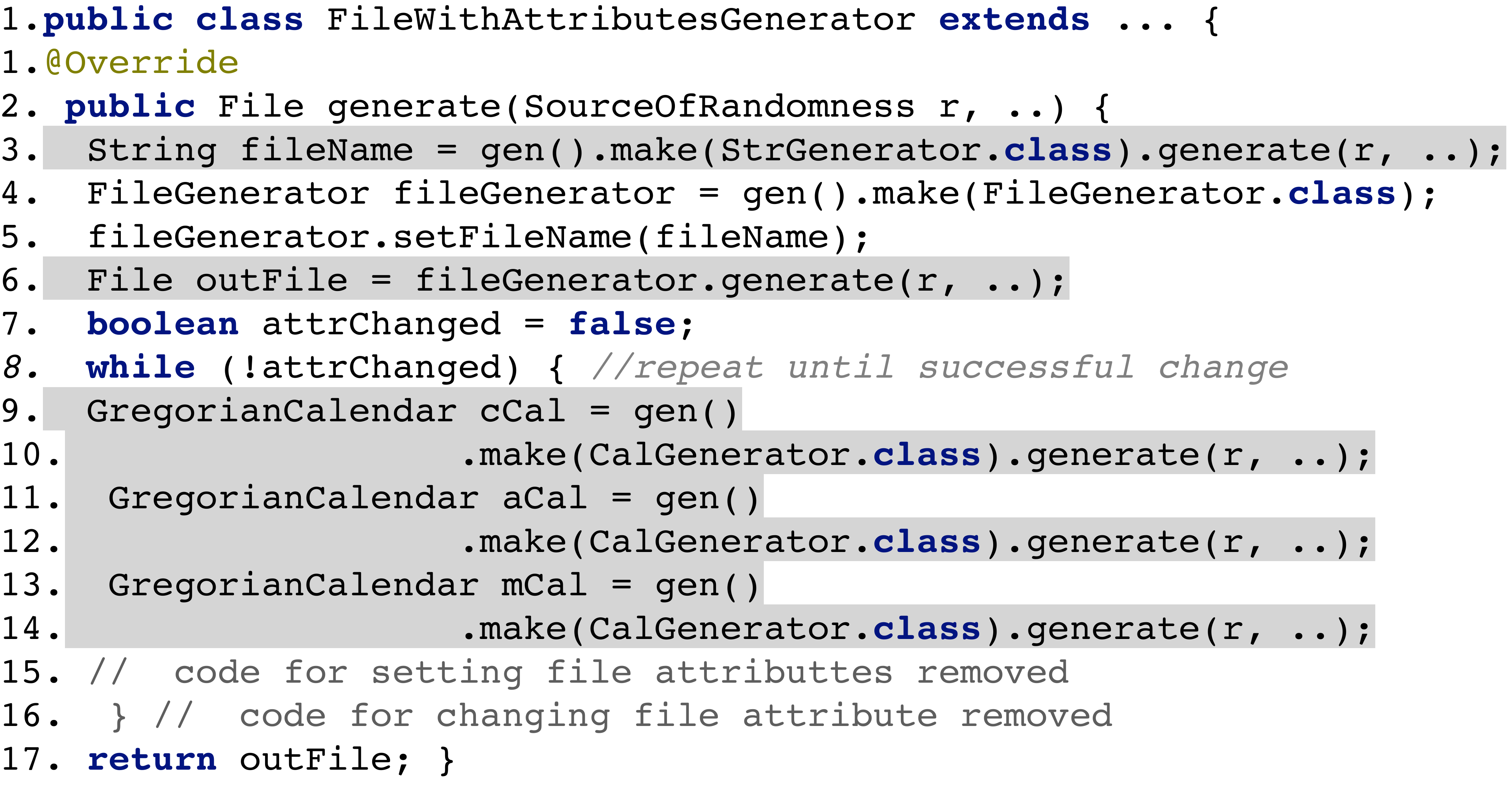}
    \caption{Snippet of the file input generator. Invocations of enclosed generators are highlighted in \hl{grey}.}
    \label{fig:filegenerator}
\end{subfigure}
\begin{subfigure}{.50\textwidth}
    \includegraphics[width=\columnwidth]{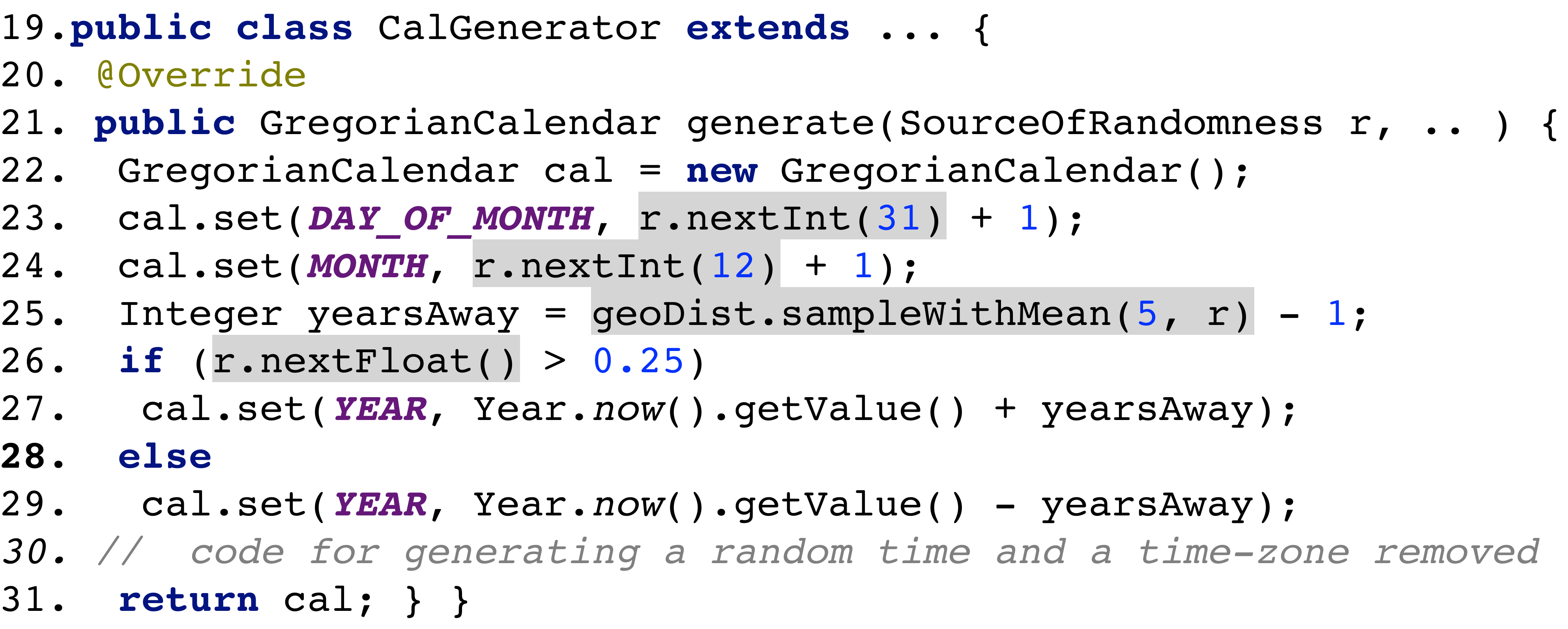}
    \caption{Snippet of the calendar generator. Random parametric choices controlled by GBF are highlighted in \hl{grey}.}
    \label{fig:calendargenerator}
\end{subfigure}
\caption{A snippet of generators}
    \label{fig:generators}

\end{figure}

\subsection{Fuzzing the Thumbnail Program with GBF}
Zest~\cite{zest} is the state-of-the-art GBF for Java programs. 
Fig.~\ref{fig:generators} shows a snippet of the generator's code written with Zest that constructs a file using a file generator {\tt FileWithAttributesGenerator} that uses, among other generators, the calendar date generator {\tt CalendarGenerator} Fig.~\ref{fig:generators} to generate dates.
Both classes define a {\tt generate} method that constructs the object of the intended type. For example, the {\tt FileWithAttributeGenerator}'s generate method starts by creating a random filename using the {\tt StrGenerator} {\bf (line 4)}, then it creates an instance of the {\tt FileGenerator} that generates the file's contents {\bf(lines 5,6)}. The remaining code generates random dates (using the {\tt CalendarGenerator}) to set the creation, the last access, and the last modification dates ({\bf lines 9-14}). The To create a calendar object, the {\tt GregorianCalendar} generator choose three random integers for its day, month and year.  The year is generated using a geometric distribution with a mean of 5, making it more likely to construct a year that is within $\pm$5 years range from the current year (25\%-75\% future and past years, respectively) ({\bf lines 26-29}).
Observe that Zest does not directly control the program input. Instead, it controls, i.e., providing an implementation for, the random selection within the input-generators represented with operations on {\tt r}, the {\tt SourceOfRandomness}, highlighted in \hl{grey}. All such selections is what we refer to as the Fuzzer Controlled Input (FCI). By controlling these random selections, the GBF can indirectly control the generation of program inputs. 
For example, the GBF fuzzer will control the random selection happening on {\bf line 5-7} in Fig.~\ref{fig:generators}. 

If one would classify the content of the FCI based on the types they are used in generating, we can see a classification that is similar to the one shown in Fig.~\ref{fig:FCILayout}. The entire FCI is used to generate an object of a {\tt File} type, which can be further decomposed to subsequences that were used to generate an object of a {\tt BufferedImage} type (the payload of the image), and the other part is used to generate the meta-data of the file ({\tt String} and {\tt GregorianCalendar} types to generate the name and the date information of the file).

The image payload occupies the majority of FIC because it is significantly larger than the metadata. As a result, the fuzzer must make far more random selections to construct the payload than the metadata. For instance, if GBF generates a 100x100 pixel image where each pixel requires red, green, and yellow color values, the payload segment in FCI would contain 30,000 data points $(100\times100\times 3)$. In contrast, the metadata segment would only contain 26-30 data points—less than 0.1\% of the total FCI. This means that with uniform mutation, fewer than 1 in 1000 mutations will target the critical metadata fields.
This imbalance in the composition of the FCI introduces a bias in mutation, that is, the random uniform mutation strategy will likely introduce modification to image's payload than its metadata. Consequently, mutations often fall outside the relevant subsequence corresponding to the file system attributes needed to achieve coverage of {\tt line 13} in Fig.~\ref{fig:motivationalExample}. 

\subsection{Fuzzing Using The Proposed Type-Based Mutation}

Making targeted mutations necessitates understanding which parts of the FCI need modification. Directing GBF fuzzers is especially challenging because of the lack of mapping between the FCI and the program's input. 
Although heavyweight techniques such as taint analysis, symbolic execution, or certain dataflow analyses can be employed to trace the propagation of the FCI within the program, this process is often complex and resource-intensive, potentially impacting the fuzzer's performance. Additionally, there are scenarios where conducting such analyses may not be feasible due to the presence of heterogeneous languages and systems. For example, it is not easy to track FCI that is going out to a file system.

In this work, we develop a new type-based mutation heuristic that directs mutations toward likely useful subsequences within the FCI, based on the types of objects they are used to generate. Our technique consists of three main steps. First, it statically identifies types that influence branch decisions in the application code. Then, while generating program inputs, it dynamically annotates subsequences within the FCI with the types they contribute to. Finally, it increases the chances of introducing mutations to subsequences whose types influence the decisions of uncovered branches.

To collect influencing types, our technique starts by identifying code targets for the fuzzing process.
A \emph{code target} is defined as the then-side or the else-side of a branching condition within the application code.
For example, the then-side and the else-side of the if-statement at {\bf line 12} in Fig.~\ref{fig:yearThumbnails} are two code targets.
To cover the then-side of {\bf line 12}, we statically compute types flowing into a branch. We call this step \emph{influencing type analysis} which employs a variation of a def-use data flow analysis. This analysis will then associate \{{\tt GegorianCalendar}, {\tt File}\} as the list of influencing types that affects the decision of that code target.

Next, during the construction of the program input, our technique maps every value within the FCI to the dynamic type of objects they are used in generating.
\begin{figure}
    \centering
    \includegraphics[width=0.98\linewidth]{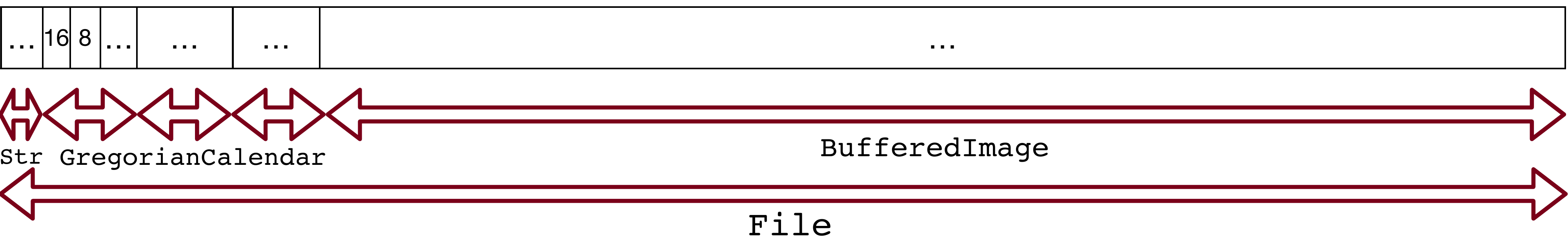}
    \caption{FCI used in generating a File from {\tt FileWithAttributesGenerator} with our extension of Type annotation}.
    \label{fig:FCILayout}
\end{figure}

For example, Fig.~\ref{fig:FCILayout} shows the FCI used in generating a {\tt File} from {\tt FileWithAttributesGenerator}, annotated by the types they are used in generating. 
Keeping such a type structure helps us direct the mutation to types that are likely to affect decisions of uncovered code targets. 
For example, suppose $v_2=8$ used on
{\bf line 23} to generate the month value. Our technique identifies that $v_2$ contributes to constructing a {\tt GregorianCalendar} object, which is itself part of constructing the {\tt File} object. These type-annotations can be computed by tracking which generator methods are active (on the callstack) when each FCI value is consumed (we detail this mechanism in Section~\ref{sec:technique}).

Finally, during mutation, our technique increases the likelihood of changing subsequences within the FCI that corresponds to influencing types of uncovered branches. The closer an influencing type is to a code target the more likely it is chosen for mutation. For example, our technique would prioritize mutating subsequences used in constructing {\tt GregorianCalendar} type, 
since it is the closest influencing type to the code target at {\bf line 12}. 

We ran the baseline fuzzer and our extension over the program in Fig.~\ref{fig:yearThumbnails}. Each run is an hour long and we repeated the experiment 20 times. Our results show that the baseline fuzzing tool was able to cover {\bf line 13} 35\% of the time (7 out of 20), while our fuzzing extension achieved the same coverage with a success rate of 70\% (14 out of 20).

%% file: technique.tex
\section{Technique}
\label{sec:technique}
\input{algorithm}

\begin{figure}
    \centering
    \includegraphics[width=0.98\linewidth]{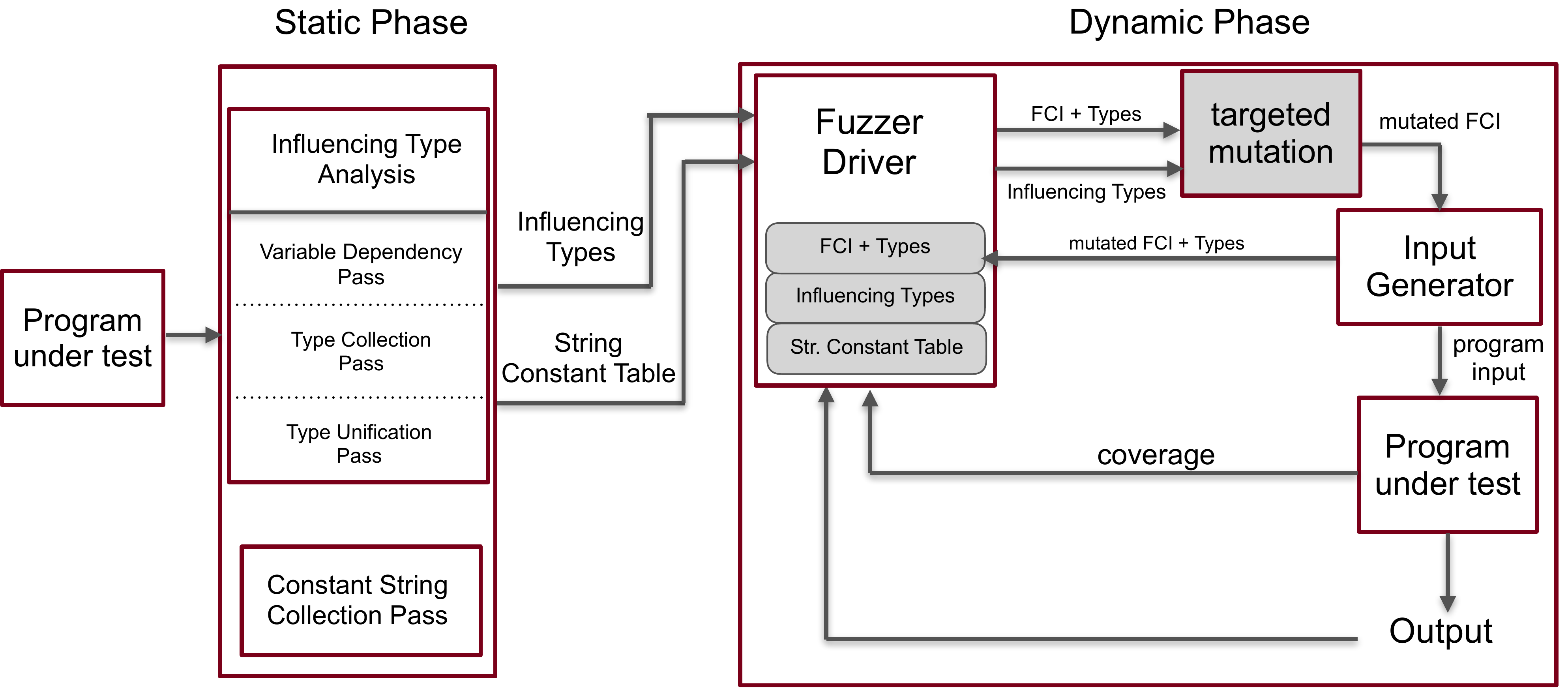}
    \caption{Overall Process of Type-Based Targeted Mutation}
    \label{fig:overallspoton}
\end{figure}
Our technique connects input generation to application code via types through three main steps (Fig.~\ref{fig:overallspoton}). First, in the static phase, we analyze application code to identify \emph{influencing types}, types whose objects flow into branch conditions. For each branch, we compute a ranked list of influencing types based on their distance from the branch decision. Second, during the dynamic fuzzing phase, we annotate each value in the FCI with the types it contributes to generating, using execution indexing to track where values are consumed. Third, during mutation, we prioritize mutating FCI segments whose types match influencing types of uncovered branches, with closer types receiving higher priority. As branches are covered, we adjust type priorities to focus on remaining coverage targets. Additionally, our static analysis collects string constants from application code to seed string generation with values likely to be meaningful in the program under test.
In the rest of this section, we present the formal algorithm for the GBF extended with our changes for the type-targeting purposes, then we discuss in detail each process within the static and the dynamic phases.

\subsection{Algorithm}
Algorithm ~\ref{alg:fuzzalgorithm} shows the main algorithm of GBF with our extension highlighted in \hl{grey}. We start by describing how GBF works, then we will introduce our extensions. 
In general, GBF's algorithm takes two inputs: an input generator $g$, and a program $p$. A generator is a function that given a FCI $I_c$, it outputs a program input $I_{p}$. Program $p$ is tested on an input from $I_p$, then terminates with either a {\scshape Failure} or a {\scshape Success} result\footnote{We assume that $p$ is always terminating. In the implementation a timeout is imposed for each test.}.
$V$ represents the set of random values while $I_c$ is the set of sequences of the fuzzer-chosen input. 
$\mathcal{S}, \mathcal{E}$ represent the set of successful and failed inputs, respectively. And finally, $\mathcal{C}_{total}$ represents the set of all code coverage.

GBF proceeds as follows. First, the set of successful inputs is initialized with a single randomly chosen FCI
{\bf(line 15)}, while the failed inputs and the total coverage sets are initialized to the empty sets {\bf(line 15)}. Then, the fuzzing loop proceeds until a specified time budget expires ({\bf lines 16-30}). 
At each step, an input $i_c$ is selected from $\mathcal{S}$, and the number of candidates to generate from $i_c$ is determined using the {\scshape numCandidates} heuristic (line 18). This function decides how many mutants will be produced from $i_c$. 
This standard GBF heuristic determines how many mutants to produce from $i_c$ based on how "interesting" it is—typically, inputs achieving more new coverage generate more children to explore the newly discovered space.
%
In each iteration, the selected input is mutated to generate $i_c'$ {\bf(line 19)}. Mutation is done by introducing random changes to random locations within $i_c$.
Next, the generator is then invoked on the mutated inputs {\bf(line 21)} to create program inputs $i_p$, which the program executes ({\bf line 22}). The fuzzer collects the result of the test and the coverage information {\bf(line 22)}. If the test fails, the program input $i_c$ is added to the set of failed inputs {\bf(lines 23-24)}. Otherwise, if new coverage was obtained, then GBF saves $i_c$ in the successful inputs  and total coverage is updated {\bf(lines 26-29)}.


 Our extension modifies the representation of the FCI. In baseline GBF, the FCI consists solely of a sequence of random values ($V$). Our extension augments this to a sequence of triples: $I_c = (V, EI, \mathcal{T}^*)^*$. Here, $\mathcal{T}$ represents the set of all possible types, $\mathcal{T}^*$ denotes the set of all finite sequences of types from $\mathcal{T}$, and $EI$ represents the set of execution indices, each capturing a unique dynamic execution location. Specifically, each element of the FCI is a triple $(v, ei, T)$ where:
\begin{itemize}
    \item $v \in V$ is a random value
    \item $ei \in EI$ is an execution index specifying the unique dynamic location within the input generators where $v$ was consumed
    \item $T \in \mathcal{T}^*$ is a sequence of types recording the return types of methods in the call stack when $v$ was generated
\end{itemize}
This representation allows us to associate each FCI value with both the types it contributes to generating and the precise location where it is consumed.

In addition to that, our extension introduces four additional data structures:
(1) A set of all code targets ($\mathcal{K}$), (2) the set of uncovered code targets ($\overline{\mathcal{C}}_{\mathcal{K}}$), (3) the map of types to their distances ($\mathcal{D}$), and (4) the map of code targets to their typing distances ($\Gamma$). 
The distance of an influencing type indicates how many def-use dependencies separate it from the branch condition. For example, if a branch directly checks a GregorianCalendar field, that type has distance 1. If the branch checks a value derived from that calendar through several intermediate computations, the calendar has higher distance. Our intuition is that types closer to the branch decision have more direct influence on the outcome, making mutations targeting those types more likely to affect coverage.

Before fuzzing starts, we analyze the program $p$ to collect influencing types together with their distances in $\Gamma$ 
({\bf line 13}), which initializes the set of uncovered code targets. Next, since a type can appear multiple times during the analysis with varying distances to different code targets, we unify their typing typing distances ({\bf line 14}) by associating a single distance to each type.

On {\bf line 20} our extension uses a type-based mutation strategy. This process takes the updated fuzzer-chosen input $i_c$, the uncovered code targets $\overline{\mathcal{C}}_{\mathcal{K}}$ and the map of typing distances  $\mathcal{D}$.
The result of this operation is a mutated input ($i_c'$), and an updated typing distance map $\mathcal{D'}$ ({\bf line 20}), which can de/prioritize types depending on whether a statically identified types was matched with a dynamic one. 
Finally, GBF proceeds by running the program on the newly constructed program input. If new coverage is achieved then the newly covered code target(s) is removed from the set uncovered code target ({\bf line 29}). We discuss each component in detail in the following sections (\ref{label:staticphase} - \ref{label:dynamicphase}).

\subsection{Static Phase}
\label{label:staticphase}

\begin{figure}
\begin{lstlisting}[numbers=none,mathescape, basicstyle=\footnotesize\ttfamily]
1.v17 = arrayload v10[v26]                
2.v19 = invokevirtual File.toPath()Path v17
3.v21 = invokestatic YearThumbnails.yearTaken(Path)I v19 
4.if(ne,v21,v3)  
\end{lstlisting}
\caption[Simplified Wala IR]{Simplified Wala IR for lines 12 in Fig.~\ref{fig:yearThumbnails}.
{\tt v10,v17,v3,} and {\tt v26} represent {\tt imageFiles, file}, the value {\tt 2008} and the length of {\tt imageFiles}}
\label{fig:ir}
\end{figure}

\begin{figure}
\begin{lstlisting}[numbers=none,mathescape, basicstyle=\footnotesize\ttfamily]
1.(Ljava/util/GregorianCalendar, 2)
2.(Ljava/nio/file/attribute/FileTime, 4)
3.(Ljava/nio/file/attribute/BasicFileAttributes, 5)
4.(Ljava/nio/file/Path, 6)
5.(Ljava/io/File, 8)
6.([Ljava/io/File, 9)
\end{lstlisting}
\caption[Simplified influencing types]{A simplified snippet of the influencing types found for the branch at {\bf line 13} in ~Fig.~\ref{fig:yearThumbnails}. Each line shows the computed influencing type and its distance from the decision.}
\label{fig:influencingTypes}
\end{figure}

In this step, our extended algorithm analyzes the \emph{application} code to find all code targets. Using this analysis we can direct the fuzzer to cover the code targets identified in this step. 
For each code target, we compute types likely to influence their decision.  An \emph{influencing type} is any type that can flow to the operands of the branching condition of the code target. 

Our analysis computes a variation of def-use relation among variables. 
We use the Wala~\cite{wala} static analysis framework to collect influencing types for this analysis, whose IR is in a static single-assignment form~\cite{ssa} representing the bytecode. 
For example, ~Fig.~\ref{fig:ir} shows a simplified Wala's IR representation of  {\bf line 12} in ~Fig.~\ref{fig:yearThumbnails}, where the {\tt imageFiles} is first loaded. This corresponds to loading elements of the array within the {\tt for} loop at {\bf line 11}. At Fig.~\ref{fig:ir}:\textbf{L2} the image file is retrieved using {\tt toPath()}, followed by invoking the application method {\tt yearTaken()} to find the year of the photo (Fig.~\ref{fig:ir}:\textbf{L3}), then finally branching on the resulting output (Fig.~\ref{fig:ir}:\textbf{L4}). Note that our analysis is not field-sensitive, for example, although the image file has three date fields, our analysis only captures the {\tt GregorianCalendar} as one of the influencing types. 
The analysis consists of three passes: a \emph{variable dependency pass}, a \emph{type collection pass}, and a \emph{type unification pass}. 

{\bf Variable dependency pass}:
In this pass, we use a flow-sensitive analysis where we create a dependency graph between variables using a variation of the traditional def-use dataflow. 
Our extension gathers variable dependencies through a depth-first traversal of the application's call graph. 
 To capture the variable dependency when a method invocation is encountered, we add more edges to the dependency graph to capture the dependencies between the actual and formal parameters of the method. Similarly, if there is a return statement within the method, then we create a dependency between the returned variable in the callee and that of the caller. 
Also, for non-static methods, we capture the def-use dependency between the variables passed to the arguments and the reference variable on which the method invocation is done.
Our analysis can identify recursive calls but visits the method only once at each encounter.
Note that, for any invocation of a non-application method, i.e., a library call or a Java standard library call, we avoid analyzing their body.
For example, the edges (use $\rightarrow$ def) in the dependency graph for this pass for ~Fig.~\ref{fig:ir} are:
\begin{align*}
   & \{{\tt v17} \rightarrow {\tt v10}, {\tt v17} \rightarrow {\tt v26}, {\tt v19} \rightarrow {\tt v17}, {\tt v21} \rightarrow {\tt v19}, & \\
   & {\tt farg({yearTaken})} \rightarrow{\tt v19},  
     {\tt v21} \rightarrow {\tt ret({yearTaken})}\} & \\
     & \cup {\tt E_{yearTaken}}&    
\end{align*}
Here, the first line captures the dependencies between def-use variables. The second line captures the dependencies encountered due to the method invocation of {\tt yearTaken()}, where {\tt farg(yearTaken)}, and {\tt ret(yearTaken)} are functions that return the variable name of the formal argument as well as the variable name of the return variable of the same method.
Observe that {\tt v21} also depends on {\tt E$_{yearTaken}$}: the set of variable-dependency edges computed from analyzing the {\tt yearTaken()} method. By contrast, the analysis of the body of {\tt toPath()} is skipped since it is not identified to be defined within the application code (within the user's outermost package name). 

Also, in this pass, we find and collect the type for each variable encountered. To find types, we use the static type signature existing in the code as well as Wala's type inference. If a variable was found to have multiple types, we keep all of them. In general, we exclude primitive types and {\tt java.lang} object types.

{\bf Type Collection Pass:}
%
In this pass, we build a dependency graph tracking how values flow through variables. Starting from a branch condition, we can trace backward through this graph to find which variables (and thus which types) influence the branch's outcome. 
%
We use distance as a proxy for influence: types closer to the branch are more likely to directly affect its outcome.
For example, ~Fig.~\ref{fig:influencingTypes} shows a simplified subset snippet of the influencing types collected for the branch at {\bf line 12} in ~Fig.~\ref{fig:yearThumbnails}. Each line is a pair of a type and a distance. Note that there might be multiple types at the same distance.  Here, we can see that the {\tt GregorianCalendar} type has distance 2 (closest to the branch), while {\tt File} has distance 8 (furthest). This ranking suggests that mutating FCI values generating {\tt GregorianCalendar} objects is more likely to affect the branch outcome than mutating values generating {\tt File} objects.
%
We exclude primitive types and their wrapper classes (Integer, Double, etc.) as influencing types because they appear ubiquitously and provide insufficient specificity for targeting. However, we handle String types specially: while we don't track them as influencing types, we collect string constants during static analysis to seed string generation.

{\bf Type Unification Pass:}
In this pass, we merge all influencing types with their distances for all code targets into a unified structure ($\mathcal{D}$). If influencing types are unique among all code targets then the unified structure will be their union. However, if an influencing type appears in multiple code targets with different distances, then the type is added to $\mathcal{D}$ with its distance in the largest code target; i.e., the code target with the largest number of influencing types. 
The intuition is that branches with more influencing types offer more mutation opportunities, so we prioritize distances that help cover these information-rich branches.

%
For example, if $(t,d_1)$ and $(t,d_2)$ are found in two code targets $k_1$, and $k_2$, then the unified structure will contain ($t,d_3$) such that $d_3=d_1,\text{ if } |k_1| > |k_2|$, otherwise $d_3=d_2$, where $|k|$ denotes to the number of influencing types associated with a code target.
Later, distances of various influencing types are updated during the dynamic fuzzing process depending on whether a static influencing type could be matched with a dynamic type. The change in distances allows the fuzzer to prioritize or de-prioritize influencing types depending on whether they were useful dynamically.

{\bf Creating Constant String Lookup Table}
In this step, we implement a simple analysis to identify magic strings, i.e., string that have specific meanings in the program under test.
Our analysis gather all reachable string constants from the call graph's entry point and within the application code. 
Later, we enable the fuzzer to alternate between selecting magic strings and generating fresh ones, when generating string values.

\subsection{Dynamic Fuzzing Phase}
\label{label:dynamicphase}
The dynamic phase uses influencing types from static analysis to guide mutations. This requires establishing a mapping between FCI values and the types they generate—a challenge because the FCI (a sequence of random choices) has no inherent connection to program inputs (structured objects).

Annotating the FCI with types alone does not guarantee that mutated values will generate the same object type. This is because introduced mutations can alter control flow within generators, meaning that values may not be consumed in the same order, which might result in creating different objects than intended. To ensure consistency, we must associate with each data point in the FCI the dynamic location within the generators where it was used.
We do this by 
using \emph{execution indexing}.
Execution indexing (EI)~\cite{orginalEiPaper,distrubtedei} of a program provides an ordered, unique representation of the dynamic trace points within the execution.
Execution indexing technique has many applications such as program alignment~\cite{programalignment}, and detecting deadlocks and concurrency failures~\cite{eideadlocks,eiconcurrency}. The main idea of execution indexing is that every program point is associated with a vector of code points that identify the unique dynamic index of a particular execution point. We use EI to ensure that values can only be reused in the same dynamic locations.
Next, we outline the process of annotating values within the FCI with types, creating their execution indices, and utilizing this information for type-targeted mutations.

\subsubsection{Using Execution Index to annotate the FCI with Types}
\label{sec:executionindex}
To annotate the FCI with types, we intercept invocations requesting a random number, as in {\tt r.nextInt()}.
Then, we collect the type of objects that are currently being constructed. 
We do that by collecting the return type of methods within callstack. For example, we intercept the execution of {\tt r.nextInt(31)} in {\bf line 23} in ~Fig.~\ref{fig:generators}. Let's assume that the entry within the FCI for this random selection is the number {\bf 9}. Then, 
we associate this value with the return types of all method invocations types in the current callstack. In this case, the value 9 will be annotated with  \{{\tt GregorianCalendar}, {\tt File}\} types; i.e., {\tt (9,\{GregorianCalendar, File\}}).
As previously noted, annotating FCI values with types alone is insufficient for targeted mutation.

In targeted mutation, the goal is to introduce localized changes to objects of a specific type while preserving the rest of the input. To prevent sequential processing of the FCI, which could disrupt this localization, we annotate each value with its unique dynamic location in the code by computing its execution index.
For example, consider the case where initially generating a {\tt fileName} required that the {\tt StrGenerator} use the first 10 values of the FCI. Suppose a mutation occurred that resulted in having {\tt StrGenerator} using only the first 5 values of the FCI. In that case, the remaining 5 values from the original FCI, while being annotated with {\tt String} types, will be used within the {\tt FileGenerator} to create an object of {\tt File} type. This is a mismatch between the object types created by the same FCI's subsequence from one test to another. 

To avoid problem
we associate an execution index with every value of the FCI. 
The main idea of the execution index is that every program point is associated with a vector of code points that identify the dynamic EI of a particular execution point. 
In our extension, we use a relaxed definition of execution indexing~\cite{baseEIPaper} implemented in Zest, where the uniqueness property among different execution indices is not guaranteed. 
More precisely, let $m_1,m_{2},.., m_n$ be the callstack when the execution of an expression $p$ is about to happen. The execution index of $ei(p)$ is defined as a vector of pairs $[l_i, q_i]$ such that ($0 < i \leq n$), where $l_i$ is the label of the statement within method $m_i$, and where  $q_i$ is the number of times the statement was invoked during the current invocation of $m_i$.

To illustrate, consider the input generator at ~Fig.~\ref{fig:generators} where a file is created and its creation date is constructed by invoking the {\tt CalGenerator} at {\bf line 13} (~Fig.~\ref{fig:generators}). If {\tt lCal} is created during the first iteration of the while loop on {\bf line 9}, then the execution index of the random choice of selecting a day of a month ({\bf line 23} in ~Fig.~\ref{fig:generators}); i.e.,  $ei(23)=[13, 1, 23, 1]$. However, if {\tt lCal} is created during the second iteration of the while loop then the execution index of the random choice of the day of a month would be $ei(23)=[13, 2, 23, 1]$. This is because the {\tt generate(r,..)} method at {\bf line 13} in this case has been invoked twice in the context of {\tt FileWithAttributesGenerator}. 

Using this approach the FCI ($i_c$) is represented as a sequence of triples of the form $i_c \in (V,EI,\mathcal{T}^*)^*$. For example, continuing our above example of creating the {\tt lCal} object ({\bf line 13}). Our extended FCI for selecting the day of the month ({\bf line 23}) will be the triple of the form: $(9, ei(23),\{GregorianCalendar, File\})$
Note that this technique is most effective when there is a separate generator method for every level of object that the generator constructs, a common coding practice.

\subsubsection{Type-Based Mutation Strategy}


During mutation, we prioritize FCI values whose types match influencing types of uncovered branches. We maintain a weighted distribution over types in $\mathcal{D}$, with weights inversely proportional to distances. We normalize these weights and perform weighted random selection to choose a target type $t$. We then search the FCI for values $(v, ei, T)$ where $t$ appears in $T$.

If matching values are found, we randomly select one and mutate a random-length subsequence of the FCI starting from that value's position. We then decrease the type's distance: $d(t) \leftarrow d(t) \times \frac{3}{4}$, increasing its future selection probability. This reinforcement reflects that finding a match confirms the type is actually generated, making it a productive mutation target.

If no matching segments exist, we increase the distance: $d(t) \leftarrow d(t) \times \frac{4}{3}$, decreasing future selection probability. This penalization reflects that the type is not currently generated, making it an unproductive target.

To maintain exploration diversity, we alternate between type-based and uniform mutation with 50\% probability. This adaptive weighting allows the fuzzer to learn which types are productive targets, focusing effort on types that both influence uncovered branches and are actually present in generated inputs.

%% file: algorithm.tex
\begin{algorithm}[t]
\SetAlFnt{\small}
\SetAlCapFnt{\small}
\SetAlCapNameFnt{\small}
\SetKwInOut{Input}{input}
\DontPrintSemicolon
\SetAlgoNoLine
\Input{A generator $g$: $I_c \rightarrow I_p$}
\Input{A program $p: I_p \rightarrow \{${\scshape Failure, Success}$\}$}
\ShowLn\Type {Set of random values $V$}
\ShowLn\Type {\hl{Set of types $\mathcal{T}$}}
\ShowLn\Type {\hl{Set of execution indexes $EI$}}
\ShowLn\Type{{Set of fuzzer-controlled inputs $I_c = V^*$} }
\ShowLn\Type{\hl{Set of fuzzer-controlled inputs $I_c = (V, EI, \mathcal{T}^*)^*$}  \textit{\quad \# replacing line 4}}
\ShowLn\KwData{Set of successful inputs $\mathcal{S} \subseteq I_c$}
\ShowLn\KwData{Set of failed inputs $\mathcal{E} \subseteq I_c$}
\ShowLn\KwData{Set of total coverages $\mathcal{C}_{total}$}
\ShowLn\KwData{\hl{Set of all code targets $\mathcal{K}$}}
\ShowLn\KwData{\hl{Set of uncovered code targets $\overline{\mathcal{C}}_{\mathcal{T}} \subseteq \mathcal{K}$}}
\ShowLn\KwData{\hl{Map of type distance $\mathcal{D}$:  $\mathcal{T}\rightarrow \mathbb{N}$}}
\ShowLn\KwData{\hl{Map of target type distance  $\Gamma: \mathcal{K} \rightarrow \mathcal{D}^*$}}
\ShowLn \hl{$\Gamma \leftarrow$ {\scshape analyze($p$)}}\\
\hl{$\overline{\mathcal{C}}_{\mathcal{K}} \leftarrow${\scshape keys}$(\Gamma)$} \quad \hl{$\mathcal{D} \leftarrow$ {\scshape unifyTypes}$(\Gamma)$}\\
$\mathcal{S} \leftarrow$ \{ RANDOM \}\quad
$\mathcal{E} \leftarrow \emptyset$ \quad
$\mathcal{C}_{total} \leftarrow \emptyset$ \\
\Repeat{given time budget expires}{
    \For{$i_c$ in $\mathcal{S}$}{
        \For{$1\leq j \leq$ {\scshape numCandidates}$(i_c)$}{
            {$i_c' \leftarrow$ {\scshape mutate} $(i_c)$} \\
            \hl{$i_c',\mathcal{D}' \leftarrow$ \mbox{\scshape mutate} $(i_c,\overline{\mathcal{C}}_{\mathcal{K}},\mathcal{D})$} \textit{\quad \# replacing line 19}\\
            $i_p \leftarrow g(i_c')$ \\
            coverage, result $\leftarrow${\scshape run}$(p,i_p)$\\
            \If{result={\scshape Failure}}
                {$\mathcal{E} \rightarrow \mathcal{E} \cup \{i_c\}$}
            \Else{
                \If{\emph{coverage} $\nsubseteq \mathcal{C}$  }
                    {$\mathcal{S} \leftarrow \mathcal{S} \cup \{i_c\}$\\
                      $\mathcal{C}_{total} \leftarrow  \mathcal{C}_{total} \cup \text{coverage}$ \\
                      \hl{$\overline{\mathcal{C}}_{\mathcal{K}} \leftarrow \overline{\mathcal{C}}_{\mathcal{K}} \smallsetminus \text{coverage}$}}
            }
        }
    }
}
{\bf return} $g(\mathcal{S}), g(\mathcal{F})$
      \caption{Generator-based fuzzer algorithm, modelled after Zest~\cite{zest}, and amended with our extension in \hl{grey}}
 \label{alg:fuzzalgorithm}
\end{algorithm}

%% file: evaluation.tex
\section{Evaluation}
\label{sec:evaluation}
Our research questions are:
\input{figures/results/benchmarks}
\begin{description}
    \item[RQ1:] Does the \tool\ achieve more coverage than GBF?
\item[RQ2:] What is the effectiveness of the type-based targeting without the constant string optimization?
\item[RQ3:] What is the overhead of the type-based targeted mutation?
\end{description}

We evaluated our technique using AWS Lambda applications. 
These applications tend to be compact in the size of user-written code while using extensive code from third-party libraries. Additionally, their inputs often possess a degree of complexity and structure, although loosely defined.
We collected our benchmark suite from GitHub. We searched for AWS Java Lambdas that are triggered by an {\tt S3EventNotification} and that use a single {\tt S3}, and/or a single {\tt DynamoDB} service. We removed applications with no branches and those that use unsupported formats.
 
Tab.~\ref{lb:serverlessbenchmarks} lists of benchmarks we used~\cite{spotongithub}, which include:
\begin{enumerate}
    \item {\tt s3-java}~\cite{s3javabench}: when an image file is inserted/updated, the lambda scales it and puts it back in the bucket. 

\item {\tt anishsana}~\cite{anishsanabench}:  when a CSV file of students' grades is inserted/updated into an {\tt S3} bucket, the lambda computes an average grade and places a record into a DynamoDB table.

\item {\tt load-historic-data}~\cite{loadhistoricbench}: when a stock ticker CSV file, the lambda collects some data from the file and creates a corresponding record onto a DynamoDB table. 

\item {\tt lambda-unzip}~\cite{lambdaunzipbench}: an archive is inserted/updated into an {\tt S3} bucket, the lambda unzips its content into the bucket and removes the archived file. 

\item {\tt csv-loader}~\cite{csvloaderbench}: when an archived CSV file is inserted/updated into an {\tt 
    S3} bucket, the lambda it extracts the CSV file; and creates a DynamoDB entry for each record within the CSV file.
    
\item {\tt nikoshen}~\cite{nikoshenbenh}:  when an object is inserted/updated into an {\tt S3} bucket, the lambda it checks whether a matching DynmoDB item exists. If not, it creates it.

\item {\tt upload-survey}~\cite{uploadsurveybench}: when an object is inserted/updated into an {\tt S3} bucket, the lambda serializes the received data, and updates the DynamoDB with their information. 
\end{enumerate}

{\bf -Implementation:} We used Zest~\cite{zest,jqf}, the state-of-the-art Java fuzzing tool to add our extension.
Zest is a GBF that allows users to write input generators. To compute coverage, Zest instruments dynamically loaded classes to collect coverage information. 
We built our extension on top of Zest; we call our tool {\bf \tool}~\cite{spotongithub}. 
Our implementation utilizes an experimental codebase within Zest that defines the FCI as a set of execution indices with values. We extended this work to amend the types of the generator's call stack to each execution index object of the FCI. 
In both tools, we added a heuristic that controls the number of mutants created from each FCI.
The heuristic ensures inputs with significant time are not overrun with excessive children. As processing time for children increases, the number of tests decreases. This heuristic embodies {\scshape numCandidates}$(i_c)$ in Alg.~\ref{alg:fuzzalgorithm}. Additionally, we utilized the Wala~\cite{wala} framework for static analysis, running it offline before executing \tool. Lastly, in \tool, mutation alternates between targeted and random mutations with a 50\% probability.

\input{tables/scpCov}
{\bf -Generators:} To fuzz test AWSLambda applications, we wrote an AWS generator package that populates AWS resources interacting with the AWSLambda application, such as S3 and DynamoDB instances. We assume that the structure of the DynamoDB table is known i.e., its primary key and range key; if one exists. 
Before every test, the entire AWS state is cleared and repopulated.
We randomly generate files of various types and add them to an S3 bucket. Supported types include XML, image, CSV, text, Java class, JavaScript, and archive files. Some of these formats already existed in Zest, others were created by the authors of this paper. These generators have a 0.003 probability of creating erroneous files, like missing or incorrect file extensions. In contrast, for DynamoDB, we randomly generate items with various column sizes (excluding primary and range keys) and populate the resource with them.

{\bf -Setup:} We used \emph{LocalStack 1.4.0}~\cite{localstack}: a local cloud software development framework used to develop, test and run AWS applications locally. We ran our experiment over a Dell Inc. Precision T3600 machine with 32 GB RAM running Ubuntu 22.04.4 LTS.
In our evaluation, we utilized Zest's Maven plugin, JQF, which instruments coverage code within dynamically loaded classes for fuzzing. 
We collected coverage information for condition coverage and method invocation coverage.
Coverage of already loaded classes (e.g., java.lang, etc.) is not included in the results of this section. Additionally, coverage for infrastructure applications like com.amazonaws and com.ibm.wala was excluded, as it is beyond the scope of this research.

\input{tables/allCov}
 
\subsection{RQ1: Does \tool\ achieve more coverage?}

To answer this question, we fuzz tested each benchmark using both Zest
and \tool for an hour with 20 repetitions. 
We categorize coverage results into two categories. (1) Application code coverage: defined as coverages of conditional branches and method invocations within the application code. (2) All code coverage: 
defined as coverages of conditional branches and method invocations within the application code as well as for those happening within all non-excluded libraries.

Tbl.~\ref{tbl:applicationcov} shows results for the application coverage, where \emph{Common} lists the common code targets found by both tools, while additional ones found by \tool, and Zest are listed in \emph{+\tool}, and \emph{+Zest}, respectively. The (*) indicates the statistical significance of the coverage, which shows that the results of 4 out of the 7 benchmarks' results are statistically significant refuting the null hypothesis that the difference in the coverage results between the two tools is only due to random variation.

Tbl.~\ref{tbl:allCov} shows results for all-code coverage, showing that the coverage of three out of the seven benchmarks is statistically significant.
In more detail, we notice that in both tables, most of the code targets were found by both tools, though \tool\ on average finds more code targets. This is mostly apparent in {\tt nikoshen} and {\tt upload-survey}. 
In {\tt nikoshen} type-targeting was useful to focus the mutation on the DynamoDB Item to create a matching and a valid item.
In {\tt upload-survey} type-targeting was useful to focus on the name of the S3 object. Both benchmarks used the constant string lookup to generate specific string constants.
In other benchmarks, we notice that type-targeting is useful for more stable results (more in the following section).

Finally, observe the correlation between application coverage and overall code coverage. Achieving higher application coverage generally leads to increased overall coverage. This is because covering a new branch in the application code often triggers the execution of additional code from third-party libraries. Thus, prioritizing application coverage is likely more efficient and leads to higher overall code coverage

\subsection{RQ2: What is the effectiveness of type-based targeting?}
\begin{figure*}
  \vspace*{-1cm}
   \begin{minipage}[t]{.5\linewidth}
   \centering
   \captionsetup{justification=centering}
    \includegraphics[scale=0.5]{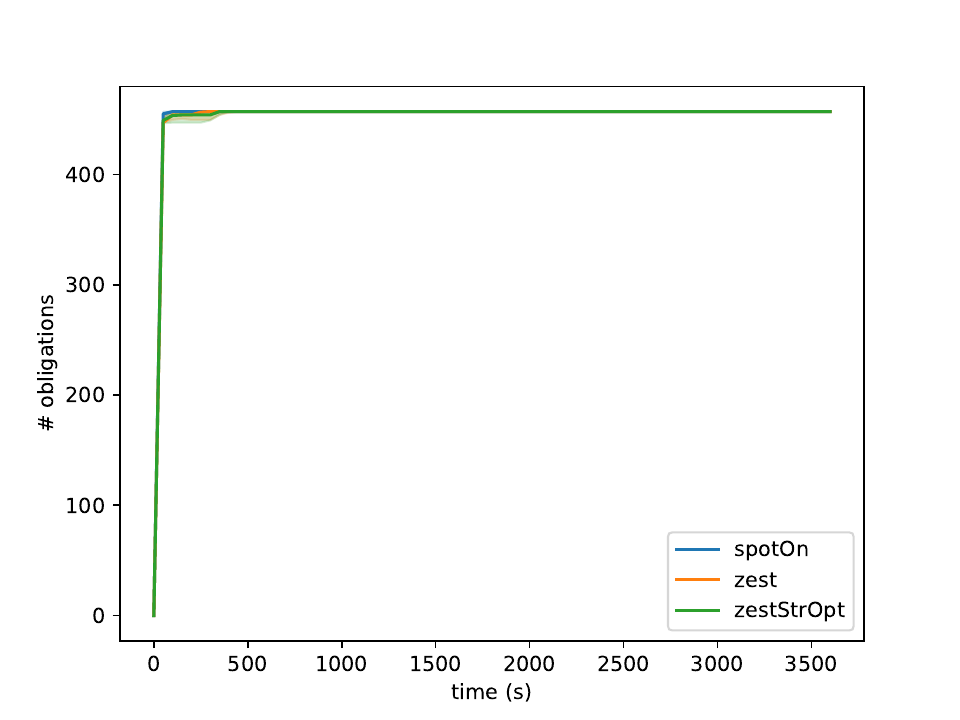}%
     \vspace*{-3mm}                                    
    \caption{Median coverage of s3-java}
    \label{fig:s3-java-coverage}
  \end{minipage}
  \begin{minipage}[t]{.5\linewidth}
  \centering
   \captionsetup{justification=centering}
    \includegraphics[scale=0.5]{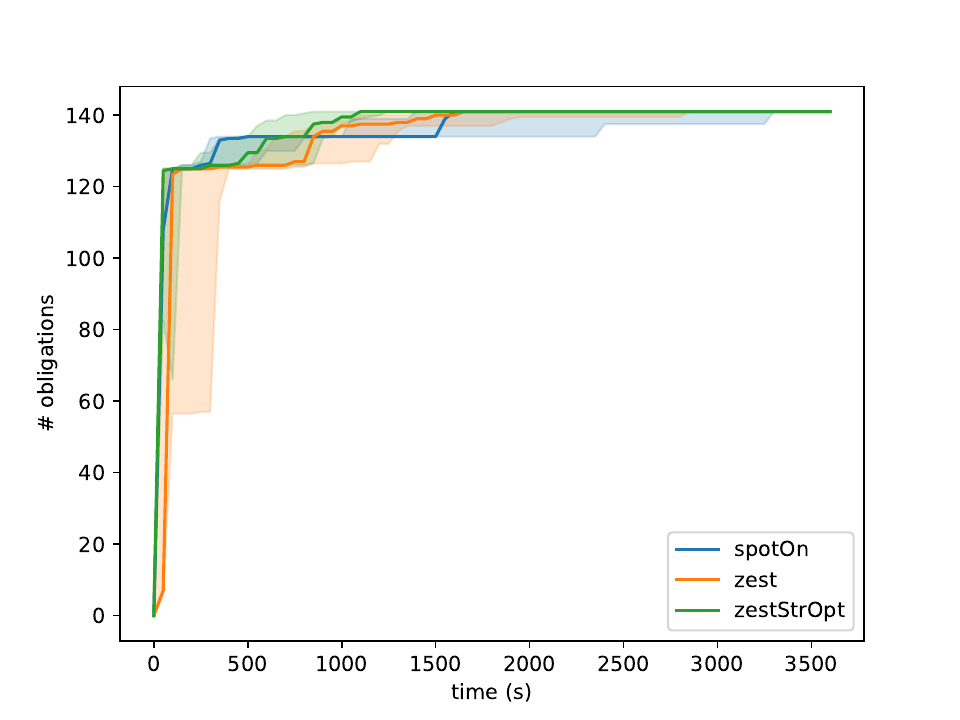}%
    \vspace*{-3mm}                                    
   \caption{Median coverage of lambda-unzip}
    \label{fig:lambda-unzip-coverage}
  \end{minipage}
  \begin{minipage}[t]{.5\linewidth}
  \centering
   \captionsetup{justification=centering}
    \includegraphics[scale=0.5]{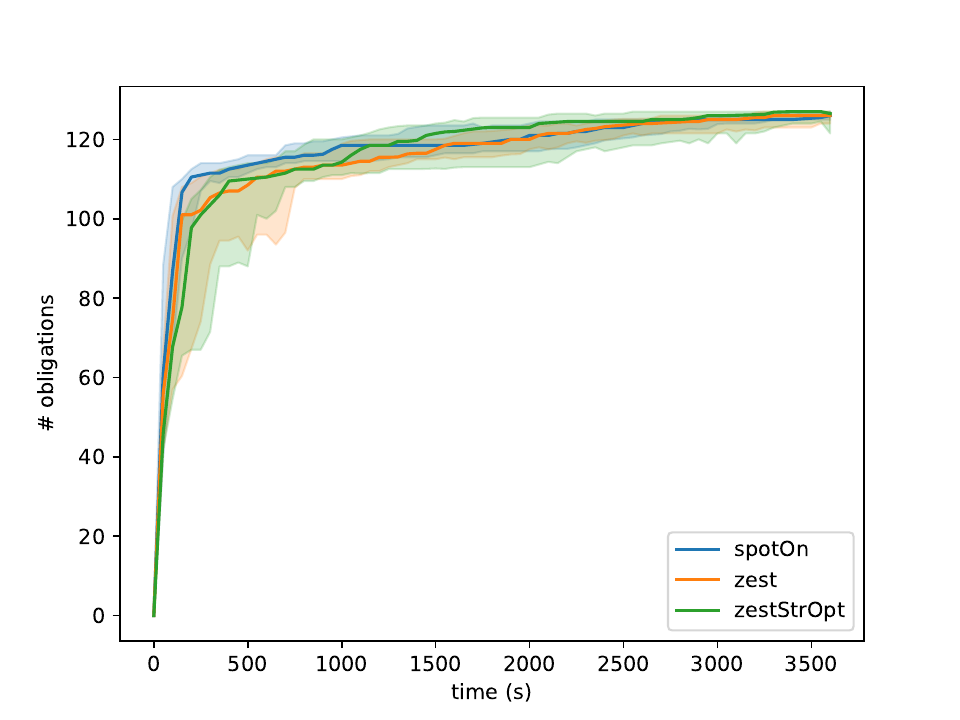}%
    \vspace*{-3mm}                                    
   \caption{Median coverage of load-historic-data}
    \label{fig:loadhistory-coverage}
  \end{minipage}%
  \begin{minipage}[t]{.5\linewidth}
  \centering
   \captionsetup{justification=centering}
    \includegraphics[scale=0.5]{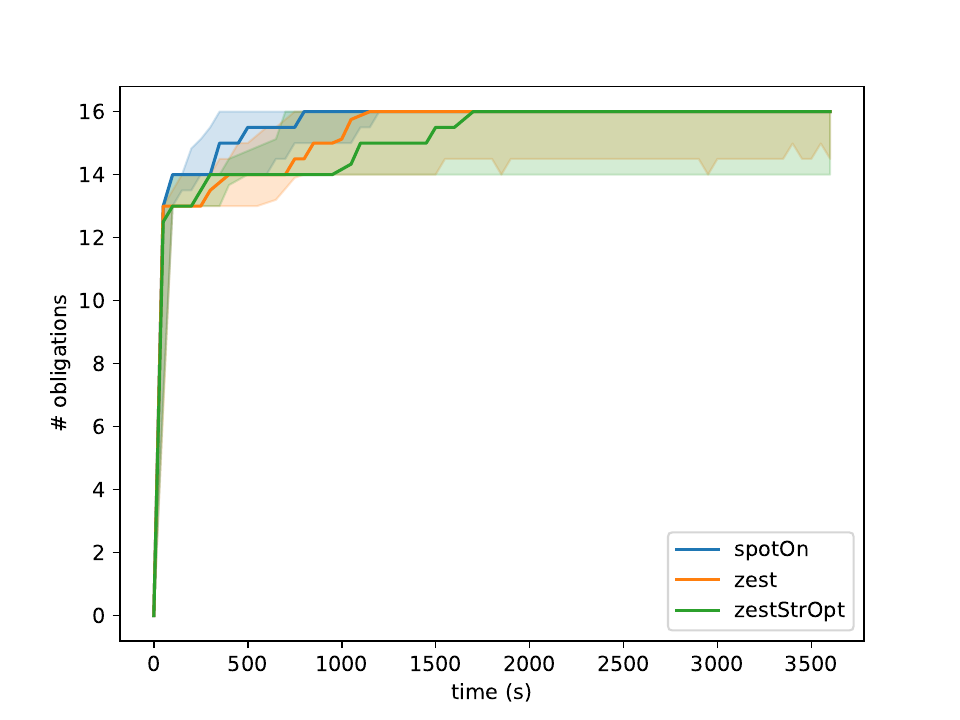}%
    \vspace*{-3mm}     
    \caption{Median coverage of anishsana}
    \label{fig:anishsana-coverage}
  \end{minipage}
  \begin{minipage}[t]{.5\linewidth}
  \centering
   \captionsetup{justification=centering}
    \includegraphics[scale=0.5]{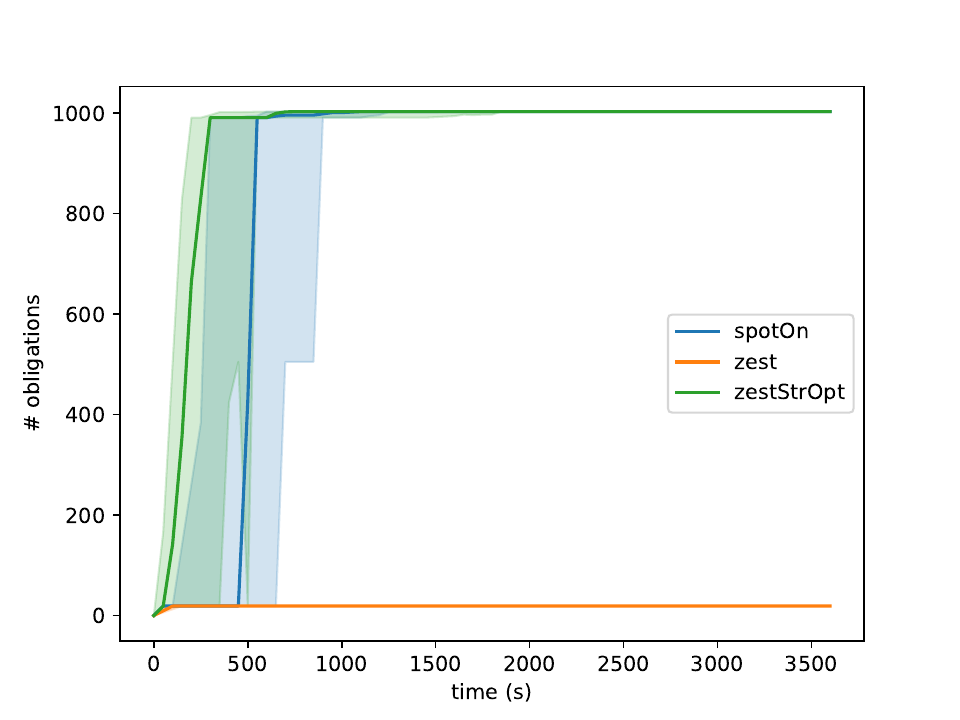}%
    \vspace*{-3mm}     
  \caption{Median coverage of nikoshen}
    \label{fig:nikoshen-serverless-coverage}
  \end{minipage}%
  \begin{minipage}[t]{.5\linewidth}
  \centering
   \captionsetup{justification=centering}
    \includegraphics[scale=0.5]{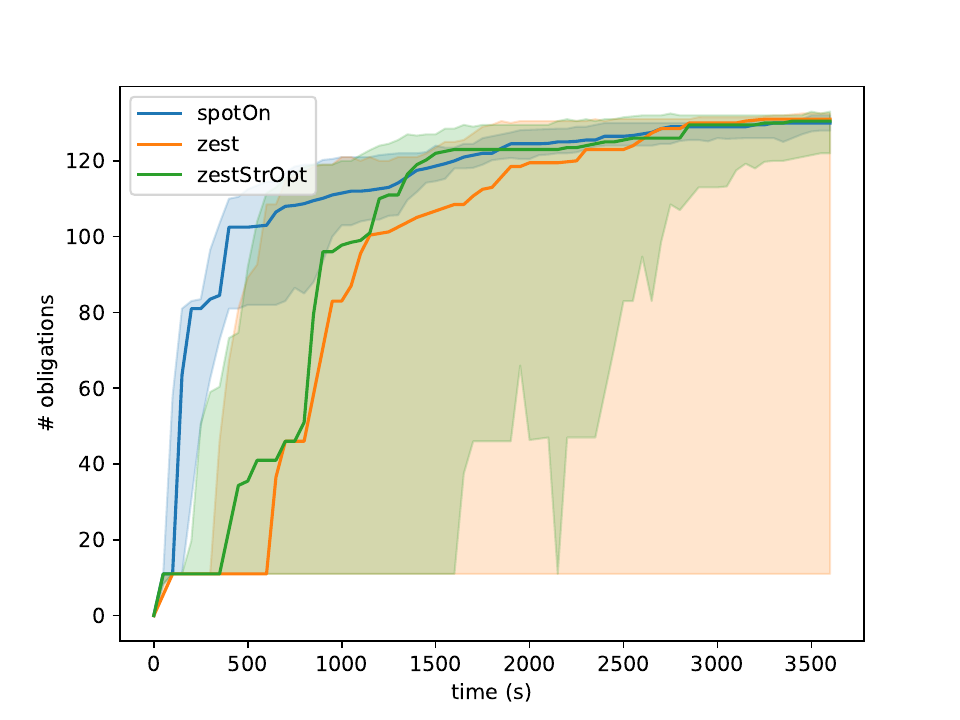}%
    \vspace*{-3mm}     
     \caption{Median coverage csv-loader}
     \vspace*{-3mm}     
    \label{fig:aws-s3-lambda-dynamodb-csv-coverage}
  \end{minipage}
  \begin{minipage}[t]{.5\linewidth}
  \centering
   \captionsetup{justification=centering}
    \includegraphics[scale=0.5]{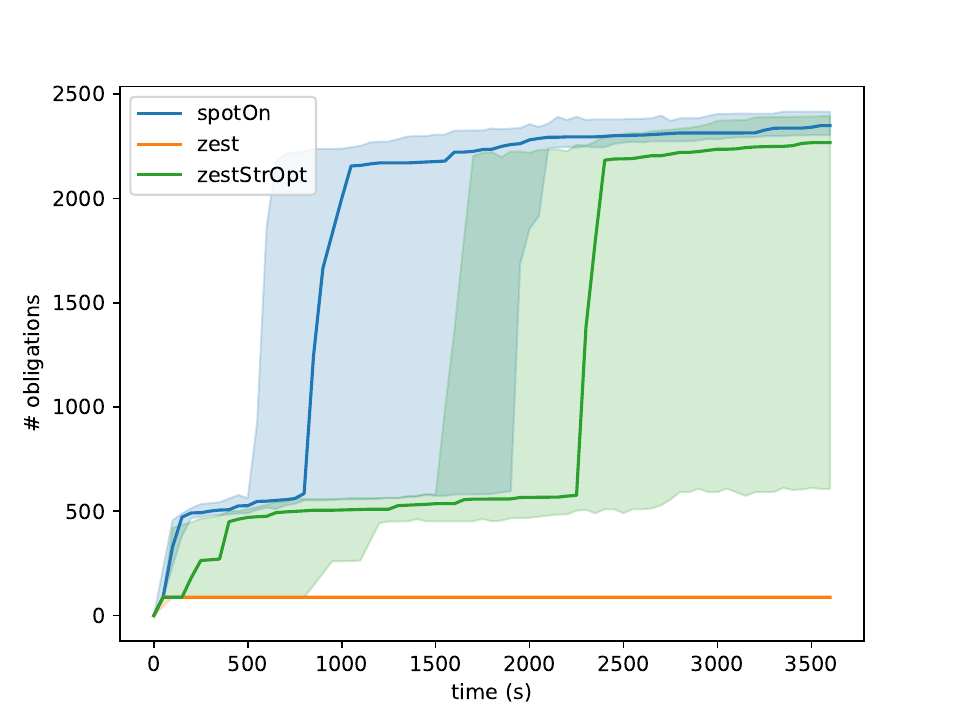}%
    \vspace*{-3mm}     
       \caption{Coverage of upload-survey}
    \label{fig:upload-survey-coverage}
    \end{minipage}
  \begin{minipage}[t]{.5\linewidth}
    \includegraphics[scale=0.5,left]{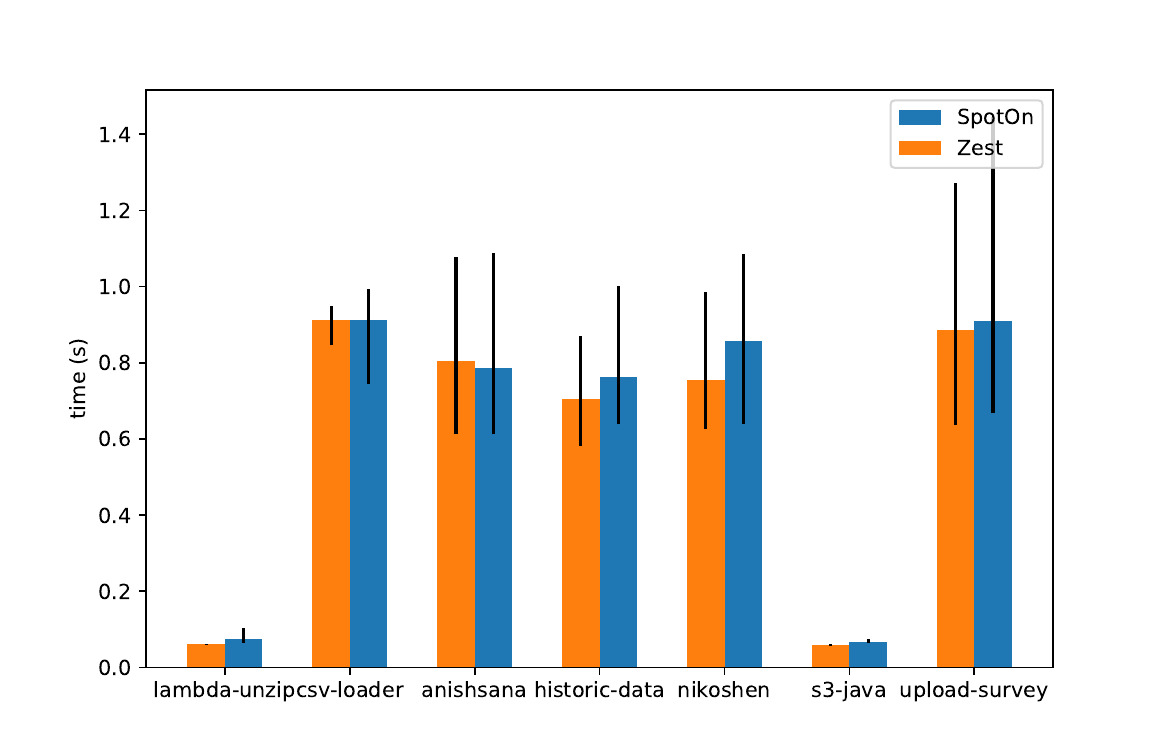}%
    \vspace*{-2mm}     
     \caption{Avg. gen. per test; errorbars between $25^{th} - 75^{th}$ percentiles}
    \label{fig:gen-cost}
  \end{minipage}
\end{figure*}

To isolate the effect of the type-based targeted mutation from the constant string optimization, we separated the later 
into a separate mode: ZestStrOpt. 
Tbl.~\ref{fig:s3-java-coverage} through Tbl.~\ref{fig:upload-survey-coverage} shows the overall coverage for Zest, ZestStrOpt, and \tool. The solid line in the graph describes the median for all 20 repetitions of the experiment. While the shaded lines describe the 95\% confidence interval. 

The first observation is that in all benchmarks, \tool\ performed either the same or better than Zest. In the case of {\tt s3-java}, neither ZestStrOpt nor \tool\ demonstrate significant coverage improvements over the baseline Zest. However, we believe this outcome inaccurately represents the effectiveness of type-based targeting, primarily due to a limitation in collecting coverage from dynamically loaded libraries. Particularly in this benchmark, the new coverage achieved by type-based mutation stemmed from pre-loaded classes, which were consequently omitted from the overall coverage collection. This limitation can be circumvented by instrumenting the code prior to the loading of any of these classes, specifically through the use of a Java Agent. Implementing this approach is a task for future work.
In {\tt lambda-unzip} (Tbl.~\ref{fig:lambda-unzip-coverage}), {\tt load-historic-data} (Tbl.~\ref{fig:loadhistory-coverage}), and {\tt anishsana} (Tbl.~\ref{fig:anishsana-coverage}), while we do not see a significant coverage difference, we can observe that \tool\ results are more reliable among the three tools. This is apparent due to its relatively smaller confidence interval. 
Similarly, we can see stable results for \tool\ in {\tt nikoshen}, {\tt csv-loader} and {\tt csv-loader}. 
We see significantly better performance for \tool\ and ZestStrOpt in {\tt nikoshen} (Tbl.~\ref{fig:nikoshen-serverless-coverage}), and in {\tt csv-loader} (Tbl.~\ref{fig:aws-s3-lambda-dynamodb-csv-coverage}). In fact, the median of the baseline Zest is at the very bottom suggesting that most of the runs were performing poorly compared to \tool\ and ZestStrOpt. The orange shading in {\tt csv-loader} suggests that a run or two from the population, Zest was able to craft the right input that achieved the desired coverage. In {\tt nikoshen} the string constant was needed to create a matching state between the DynamoDB item and the S3 object. In {\tt csv-loader} the string constant was needed to create an S3 object with a particular name. In both benchmarks, the type-based targeted mutation was useful to speed up the coverage by focusing the mutation on the substructure with the matching influencing type. In general, we can conclude that \tool\ has the most reliable performance in coverage finding among the three extensions. 

\subsection{RQ3: What is the overhead of \tool?}

To identify the performance bottleneck for \tool, we divided the fuzzing operations into four processes: 
\begin{enumerate}
    \item mutation: the process of mutating subsequences of FCI, 
    \item generation: the process of constructing an input by running the input generators on the FCI, 
    \item testing: the process of running the program under test with the input generated from the generation step, 
    \item handling: the process of handling test result, i.e., new coverage, success or failure.
\end{enumerate}

Our primary experiment's data revealed that the combined impact of mutation and handling processes accounts for no more than 3\% of the total execution time. To assess the effect of the generation and testing steps, we conducted a controlled experiment in which \tool\ managed two distinct FCI types: a linear FCI to mimic Zest's operations and an EI-enabled type-annotated FCI. We utilized the linear FCI for the generation phase and the annotated FCI for the mutation process. This setup ensures that both Zest and \tool\ produce the same inputs, but \tool\ additionally creates and maintains the annotated FCI.
We ran both tools using this configuration for a fixed number of tests. For each benchmark, we used three-quarters of the tests from the one-hour time-out experiment. The following results report the averages of the 10 runs.

Tbl.~\ref{fig:gen-cost} shows the average generation time per test. We notice that the average generation cost of \tool\ is not too far off from Zest's.
This overhead is due to computing and maintaining the annotated FCI. More precisely, during the generation \tool\ intercept, every method invocation to keep track of EI counters used in their creation. Then, when a {\tt java.util.Random} method is about to execute, \tool\ creates the EI instance and looks up its corresponding value within the FCI. If found, the generation uses the present value, otherwise a fresh value is created and the triple of the EI object, its annotated type, the fresh value is added to the FCI. 

Also, we measured the usage of the Java heap during experiments. We noticed that both Zest and \tool\ have a high rate of heap allocations-deallocations forming a sawtooth-like shape for heap usage over time. 
Overall, we observed that \tool's memory overhead ranges from a minimum of 0.82x to a maximum of 4.62x, with an average overhead of 3.29x when compared to Zest's.

\section{Threat to Validity}
The effectiveness of the GBF depends on its generators, and so our results are dependent on the generators that we have constructed. We created these generators because there are no previously written generators for the types we used in our experiments, i.e., AWS-related types. 
In general, the results can be dependent on \emph{what} the generators produce and \emph{how} they are defined. 
The former is an inherent limitation in GBF, and we addressed it by designing generators to produce random inputs for all decision-relevant data typically used in programs while keeping others, like {\tt eventTime} and {\tt eventVersion} in {\tt S3Event}, constant.

The latter limitation can arise where the linear reuse of values in the FCI may unintentionally alter the structure of created objects (see~\cite{havoc}), potentially impacting Zest's results. This limitation is mitigated by designing generators that construct objects in a logical order without relying on any specific program under test.

%% file: figures/results/benchmarks.tex
\begin{table}[t]
\centering
\caption{List of AWS lambda applications obtained from GitHub with the number of enclosed lines of code (loc)}
\begin{tabular}{|l|l|l|}
\hline
{\bf Serial}    & {\bf benchmark}    & {\bf loc}         \\ \hline
1      & s3-java    & 174        \\ \hline
2 & anishsana      &  101        \\ \hline
3 & load-historic-data   & 182     \\ \hline
4 & lambda-unzip    & 170           \\ \hline
5 & csv-loader   & 407               \\ \hline
6 & nikoshen      & 131              \\ \hline
7 & upload-survey    & 16,661    \\ \hline
\end{tabular}
\label{lb:serverlessbenchmarks}
\end{table}

%% file: tables/scpCov.tex
\begin{table}[]
\centering
\caption{Average application coverage, (+\tool): only found only by \tool, (+Zest): only found by Zest. ($^*$) indicates statically significant results (p-value = 0.05)}
\label{tbl:applicationcov}
\begin{tabular}{|l|r|r|r|}
\hline
\textbf{benchmark} & \multicolumn{1}{l|}{\textbf{common}} & \multicolumn{1}{l|}{\textbf{+\tool}} & \multicolumn{1}{l|}{\textbf{+Zest}} \\ \hline
lambda-unzip                  & 21.2           & 1.80                  & 0                     \\ \hline
csv-loader$^*$          & 13.45           & 4.45                  & 0.7                   \\ \hline
anishsana          & 8.45            & 0.50                   & 0.05                   \\ \hline
load-historic-data$^*$ & 13.15           & 0.80                   & 0.05                     \\ \hline
nikoshen$^*$           & 12.60           & 2.9                  & 0                     \\ \hline
s3-java            & 16.65           & 0.35                   & 0                     \\ \hline
upload-survey$^*$       & 10.00           & 6.10                  & 0                     \\ \hline
\end{tabular}
\end{table}

%% file: tables/allCov.tex
\begin{table}[]
\centering
\caption{Average of \emph{all} (third-pary-included) coverage, (+\tool): only found by \tool\, (+Zest): only found by Zest. ($^*$) indicates statically significant results (p-value = 0.05)}
\label{tbl:allCov}
\begin{tabular}{|l|r|r|r|}
\hline
\textbf{benchmark} & \multicolumn{1}{l|}{\textbf{common}} & \multicolumn{1}{l|}{\textbf{+\tool}} & \multicolumn{1}{l|}{\textbf{+Zest}} \\ \hline
lambda-unzip       & 126.8          & 14.15                 & 1.05                     \\ \hline
csv-loader         & 86.3           & 42.9                 & 3.45                  \\ \hline
anishsana          & 15.45            & 0.5                   & 0.05                   \\ \hline
load-historic-data  & 119.5          & 7.05                 & 1.6                  \\ \hline
nikoshen$^*$          & 218.1          & 702.55                & 0                     \\ \hline
s3-java            & 485.5          & 0.50                   & 0                     \\ \hline
upload-survey$^*$       & 88           & 2350.55               & 0                     \\ \hline
\end{tabular}
\end{table}

%% file: related-work.tex
\section{Related Work}
\label{sec:related-work}

Coverage-guided fuzzing (CGF)~\cite{aflsite, aflnet, aflfast, aflgo,tensorfuzz, coverageguidedtracing,aflsmart}
    is a random testing technique with lightweight instrumentation for coverage computation. Using the coverage information, fuzzing tools can then make informed decisions about the choice of interesting input likely to achieve new coverage if mutated. CGF revealed many bugs in widely used/tested programs, such as in Clang, OpenSSH, JavaScriptCore, LibreOffice, Python, SQLite, Google closure-compiler, JDK, Mozilla, and BCEL to mention a few~\cite{aflsite, libfuzzertrofies, jqfgithub}. 

    However, as CGF lacks information about the input structure, its effectiveness can be negatively affected.
    Grammar-based fuzzers such as CSmith~\cite{CSmith}, jsfunfuzz~\cite{jsfunfuzz},  Grammarinator~\cite{grammarinator}, and others~\cite{SemanticsExplorationforBrowserFuzzing,NAUTILUS,EvolutionaryGrammarBasedFuzzing,GramFuzz,Superion}
    use a declarative form of the input that defines input\rq s grammar.
%
    %
    Some works used the existence of a test corpus to craft new inputs from existing tests~\cite{Superion,skyfire,langfuzz,die}. Most of this work relied on understanding grammers and/or features of existing corpus and then use structural mutation to generate new inputs. 

Generator-based fuzzers (GBF) utilize programmatically defined generators to create structured inputs. 
While both grammar-based fuzzers~\cite{SemanticsExplorationforBrowserFuzzing,Generatinghighlystructuredinput,GramFuzz} and generator based fuzzers~\cite{zest,zeugma,BeDivFuzz,RLCheck} are more likely to trigger semantic behavior within the program under test, due to their knowledge of the structure of the input, we view them as alternative techniques to create a structure, semantically correct input. However, we observe that the latter naturally imposing language-type structure, which we leverage to focus the generation on likely useful substructures within the input.
     Padhye el.~\cite{zest} introduced generator-based fuzzers by extending generator-based testing~\cite{quickcheck} to use random input generators. 
     In our work, we extend Padhye el.\rq s work to enable type-based targeted mutation.

    In CGF, obtaining useful seeds has been an active area of research~\cite{aflfast,aflgo,mu2,inputswithnaturallanguage}, where some work focused on developing heuristics for picking seeds, and others focused on finding values to create good seeds.
    %
    %
For example, Marcel Bohme et al.\cite{aflfast} used a Markov chain to assign energy to seeds, prioritizing inputs that exercise low-frequency paths for better coverage. AFLGo\cite{aflgo} computes the closeness of a seed to targeted code locations and gradually assigns more energy to seeds closer to the target. While these approaches focus on input selection heuristics, our work concentrates on mutating inputs based on their types. $\mu^2$\cite{mu2} controlled seed selection by computing a mutation score for each seed, while Pythia\cite{Pythia} used a learning-based technique to identify useful mutations. FairFuzz~\cite{fairfuzz} computes rare test targets and freezes parts of seeds that contribute to reaching these targets, mutating the remaining parts.
%
%
    Other tools\cite{grammarbasedwhitetboxfuzzing, driller,confetti} use a form of symbolic execution to compute useful values for the input. The key advantage of this technique is that, unlike the heuristic-based approach, symbolic execution can turn the search for new inputs into a constraint set. When these constraints are successfully solved, they yield a new input that effectively covers the intended targets. The cost, however, is that these tools are usually slow due to the overhead of constructing and solving constraints.

     The work we present in this paper falls into the first category of heuristics for picking seeds, which tries to craft a likely useful input using type-based targeted mutation. In contrast with FairFuzz, our work finds the places where mutations are needed, rather than the places that should be avoided. 
     More recent work in GBF~\cite{BeDivFuzz,havoc} highlighted the importance of identifying the structure of the input to increase the effectiveness of the mutation. However these techniques focused on orthogonal aspects of mutation, namely,
     whether change affects the object structure~\cite{BeDivFuzz}, or side-effects of mutations on subsequent choices~\cite{havoc}.

%% file: conclusion.tex
\section{Future work}
\label{sec:futurework}
One limitation of this approach is its ineffectiveness with recursive types,
as the entire FCI will be made up of a few repeated types making the type-based targetting less effective
We plan to extend this work to support effective targeting for objects of recursive types.

We focused on testing serverless applications, as they are ideal candidates for our technique. 
We plan to evaluate our technique on traditional programs. Unfortunately, existing traditional programs rely on using primitive or recursive types, unsuitable for
our technique.
Thus, we will seek suitable benchmarks for traditional programs to assess our approach.


Finally, we plan to investigate the automatic generation of the generators, a common hurdle for the usability of GBF.

\section{Conclusion}
\label{sec:conclusion}

We introduced a novel type-based mutation heuristic with constant string lookup for generator-based fuzzers. 
Results show improvements in application coverage of an average
of 18\% over the baseline, and improvements in coverage including third-party code of 43\% when the set of covered
libraries remained the same, and even more so when our approach covers
additional libraries.